# THE FEASIBILITY OF SHADING THE GREENHOUSE WITH DUST CLOUDS AT THE STABLE LUNAR LAGRANGE POINTS

**CURTIS STRUCK**
*Dept. of Physics and Astronomy, Iowa State University, Ames, IA 50014 USA*
Email: curt@iastate.edu

There are many indications that anthropogenic global warming poses a serious threat to our civilization and its ecological support systems. Ideally this problem will be overcome by reducing greenhouse gas emissions. Various space-based methods, including large-scale solar shades, diffusers or atmospheric pollutants, have been considered to reduce the solar constant (input flux) and the warming in case emissions reductions are not achieved in a timely way. Here it is pointed out that proposed technologies for near-Earth orbiting comet deflection, suggest a different kind of space-based solar shade. This shade would be made up of micron-sized dust particles derived from comet fragments or lunar mining, and positioned in orbits near the triangular Lagrange points of the Earth-Moon system. Solar radiation pressure can render such orbits unstable, but a class of nearly resonant, and long-lived orbits is shown to exist, though the phase space volume of such orbits depends on dust grain size. Advantages and disadvantages of this scheme relative to others are considered.

## 1. INTRODUCTION

Ever more evidence of the effects of global warming and its anthropogenic drivers appears frequently in the scientific literature and the press. This warming threatens our natural support systems and could undermine our civilization [1]. Both short-term amelioration strategies and long-term cures depend on reducing the emission of greenhouse gas via the development of alternate energy, transportation and heating/cooling technologies. Unfortunately, it will take decades or more to put the needed technologies into service [2], even when they are available. In the meantime, forces such as population growth and global economic development exacerbate the problem. Thus, still longer times will be needed to reduce the abundances of greenhouse gases. In these dire circumstances, solutions other than the reversal of the individual driving forces deserve consideration, at least if they can be shown to have no comparably deleterious consequences. A number of external solutions with global impacts have been considered in the last few decades, including: injecting condensation nuclei into the Earth's atmosphere, shading the Earth with a particle ring, or shading the Earth with a shield or lens at the Lagrange point between the Earth and the Sun (see references 3, 4, and references therein). Here a novel astronomical solution is considered – reducing the solar insolation (energy input) for a period of decades or more by distributing dust into an obscuring cloud in a torus around the lunar orbit, or around the L4, L5 Lagrange points. The latter are each located at the vertex of an equilateral triangle with the Earth and Moon at the other vertices.

## 2. A DUST SHADE

For example, consider the feasibility of shading by a dust cloud centered on one of the triangular Lagrange points. Assume the cloud is roughly spherical and homogeneous, with a diameter such that it subtends 10° on the sky as viewed from Earth. Since the lunar orbit is inclined by about 5° from the ecliptic plane a cloud of this size would eclipse the Sun for at least some part of every month regardless of the location of the nodes (intersections between the two planes). At a distance of $3.8 \times 10^8$ m the cloud diameter is about $6.6 \times 10^7$ m. Suppose further that the cloud consists of roughly spherical grains with a size of about 20 microns, or a radius of $R_g \approx 10$ microns.

We will consider several possible sources for the dust grains below, including captured comets, so it is worthwhile to compare the cloud properties to those of comet materials. We have much yet to learn about the size distribution of grains sublimed from comets. However, this estimate is of the right order of magnitude. It is also the typical grain size detected by the Stardust mission in the tail of comet Wild [5], and by the Deep Impact mission to comet Tempel 1 [6]. The cross section for scattering optical and near-infrared sunlight by such grains would be approximately the geometric value, $\sigma_g = \pi R_g^2$.

To accomplish the shading, suppose that we want an optical depth of $\tau = 1.0$ for a line-of-sight piercing the cloud at a distance of 90% the cloud radius from the center (corresponding to a sky angle of about 4.5° from the center). This optical depth reduces the solar flux by a factor of $e^{-1} = 0.368$. At the cloud center the optical depth is 4.7 and the flux reduction factor is 0.0090, very dark indeed. Then the path-length of the $\tau = 1.0$ beam through the cloud would be $l = 2.9 \times 10^7$ m. Using the relation $\tau = n\sigma_g l = 1$, we find a grain density of $n = (\sigma_g l)^{-1} \approx 110$ m$^{-3}$. If the internal density of the grains themselves is about 3000 kg m$^{-3}$, then the grain mass is $1.2 \times 10^{-11}$ kg. for spherical grains, or a factor of a few less for spinning disk-like grains of comparable cross section. The total mass of (spherical) grains in the cloud is about $2.1 \times 10^{14}$ kg.

A spherical, constant density comet of typical diameter 10 km, with a typical density, has a mass of about,

$$M_c = 3.1 \times 10^{14} \left(\frac{\rho}{600 \, kg \, m^{-3}}\right)\left(\frac{R_c}{5 \, km}\right)^3 kg. \quad (1)$$

Since dust makes up a large fraction of of the cometary material (see [5], with recent results from the Deep Impact mission in [8]), a single comet of this size would have nearly enough material to supply the hypothetical cloud.

Given that the shadow of the completely eclipsed Sun makes such a small track across the Earth's surface, one might question whether the dust clouds considered above really shadow the whole Earth. Classical geometric arguments using similar triangles show that if the Moon was about three times its size (1.5° on the sky) then the Earth would be completely in shadow when the Earth, Moon, and Sun are lined up. Thus, whenever the sun is behind the dust cloud and the shortest angular distance from solar limb to the effective cloud edge is more than 0.5° the Earth will be completely shadowed.

However, the question remains – would such a cloud provide enough shade to mitigate the effects of global warming? For a simple estimate, suppose that the cloud blocks most of the solar radiation for a fraction of the time equal to 5°/360° = 0.014, or 1.4%. (Depending on the position of the lunar nodes this figure could range from 0 to 2.8%.) Even when the Sunlight passes through an optical depth of 1.0, a good deal of light will get through, but the above percentage could be roughly doubled with clouds at both L4 and L5, so we continue to use it as a rough estimate.

Using the ideal thermal radiation approximation $F \propto T^4$, we find that if the Earth's mean surface temperature is about 290K [7], and if we want reduce its equilibrium value by 1.0 K, we must reduce the incoming flux by 1.4%. In the next century global warming is expected to be order several Kelvins, so a reduction of order 1.0 K would be significant. In fact, a smaller reduction would still be very helpful as discussed in Sec. 3.1.

It is remarkable that these estimates of the amount of material needed relative to a typical comet, and the likely efficacy of the cloud for mitigating global warming, work out this well and comparable to proposed artificial shades at the Earth/Sun L1 point [3, 4]. Given that, it is interesting to consider the stability and the lifetime of the cloud. It would be desirable for this lifetime to be longer than a few decades to allow time for anthropogenic greenhouse gas sources to be reduced. At the same time, it would be desirable to have a lifetime that is less than centuries. In the classical restricted three-body system particles of much lower mass than the two primary bodies can remain indefinitely in so-called tadpole orbits in the stable basins around the L4, L5 points [7, 9].

However, since the Sun's tidal force is about 44% that of the Moon's, it could have a significant effect on the system, which varies at any point over the course of the lunar month. In celestial mechanics the full system is often referred to as a restricted four-body system, and it is well known that the restricted three-body theory of the Lagrangian stability basins is no longer accurate. Nonetheless, station-keeping for an artificial satellite near the L4, L5 points would require relatively little energy [10]. Beyond this detailed numerical studies have shown that particles as far as 5 x 10$^7$ m from L5 can remain in orbit around it for time spans of up to a hundred years [11-13].

Grain motions will not be purely orbital, since collisions are likely. The grain cross section for both photo-absorption and for collisions with other grains is the geometric cross section. Therefore, the mean free path for grain collisions will be about the size of the cloud. The random velocities of grains sublimed from cometary surfaces are small [7], so grain collision times are probably much longer than the orbital time scale (a lunar month). With low relative velocities, grain collisions don't generally result in large orbital changes, but rather longer diffusive changes result from many collisions.

The forces that create comet tails, solar radiation pressure and the solar wind, would also act these dust clouds. A preliminary estimate of the timescale for radiation pressure to impart a velocity impulse Δv to a grain is,

$$T_{rad} = \frac{\Delta v}{a_{rad}} = 76 \left(\frac{\rho_{gr}}{3000\,kg\,m^{-3}}\right)\left(\frac{R_{gr}}{10\,\mu m}\right)\left(\frac{\Delta v}{1.0\,km\,s^{-1}}\right) \text{days},$$

$$\text{using, } a_{rad} = \frac{L_{sun}}{4\pi c d_{sun}^2 \rho_{gr} R_{gr}}, \quad (2)$$

where $c$ is the speed of light, $d_{sun}$ is the distance to the sun, $L_{sun}$ is the solar luminosity, subscript $gr$ refers to grain quantities, and we assume that the absorption cross section equals the geometric cross section for a simple estimate. This time is somewhat longer than two lunar months. If it was shorter than a lunar month, then the grain would be pushed directly out. If it were much longer than a lunar month, then through the lunar cycle the pressure would alternately push the grain ahead in its orbit, and pull it back, with no net effect. In the present case, a significant fraction of the grains could be distributed around the whole L4, L5 and lunar orbit within a few lunar months, and pushed out of the system on a somewhat longer time scale. Yet other effects, such as self-shading, complicate the picture. We will take this issue up again in section 4 with the help of some simple numerical models (also see 14).

To summarize the discussion to this point, a dust cloud derived from the material of one comet may persist for some time around a triangular Lagrange point of the Earth-Moon system, and could provide significant insolation reduction to partially offset global warming during that time. Unfortunately, nature is unlikely to deliver the comet when needed, and the large mass of comets makes it impossible to move one into place with current spacecraft. (However, if the so-called Kordylewski clouds exist, then nature has placed some material at this location, see 15, 16). Nonetheless, this task may not be beyond the reach of our technology.

## 3. OBTAINING MATERIALS FOR THE DUST SHADE

In the previous section we pointed out that a comet could provide the material for a Lagrange point dust shade to reduce solar insolation. We will consider the feasibility of this in the following subsection. However, this is not the only means of constructing a global shade. We will briefly review the possibility of using lunar material in Section 3.2, and consider other shading possibilities in Section 5.

### 3.1 Constructing the Shade with Captured Comets

Various methods for comet deflection have been considered in some detail in recent years, to prevent catastrophic collisions. The recently proposed gravitational tractor of Lu and Love [17] is a simple, cheap, and thus, very attractive solution for this purpose. However, the low thrusts involved make this a less likely means of actually capturing a comet. Inevitably, substantially changing the orbit of a comet of significant mass (rather than slightly deflecting it) on decadal time scales would involve substantial orbital velocity changes, and large thrusts would be needed to generate such impulses. Other methods, ranging from nuclear engines to solar sails can generate much larger thrusts [18-20]. Probably the best method for the present purposes is the solar collector or "telescope tugboat" of Melosh *et al.* [18, also see 20], which uses concentrated solar energy to create propulsive sublimation on the surface of a comet.

These authors argue that a spacecraft with a curved mirror of up to 10 km in diameter could be constructed and sent to an asteroid or comet with current technologies. (However, difficulties, such as degradation from the ablated cometary material remain to be fully resolved.) They estimate that this craft could generate a thrust on the comet of about $1.2 \times 10^7$ N, with water molecules expelled at about 1 km s$^{-1}$ from the surface. (The momentum thrust to input energy ratio is high for water.) This would yield an acceleration of $6.1 \times 10^{-7}$ m s$^{-2}$ for a comet of 2 km in radius, and velocity change of,

$$\Delta v = 19 \left(\frac{t}{1\,yr.}\right)\left(\frac{R_c}{2\,km}\right)^{-3} m\,s^{-1}. \qquad (3)$$

This is still a small quantity compared to the 1 km s$^{-1}$ orbital velocity of the Moon around the Earth, or the 30 km s$^{-1}$ orbital velocity of the Earth around the Sun. It would require a velocity change of about 10 km s$^{-1}$ to capture a comet with maximum and minimum orbital radii of about 1 and 5 A.U., respectively (typical of near Earth orbiting comets). With the energy supplied by half a dozen to a dozen tugboats and a timescale of a decade, this net impulse could be achieved according to equation (3).

Note that it would take a much longer time to achieve it with the $R_c$ = 5 km comet discussed earlier. Thus, we would have to settle for a smaller obscuring cloud (2-4° versus 10°), fetch more small comets, or use the material more efficiently. An example of the latter would be to spin the captured material around an axis pointing toward the Earth, so that it diffuses outward in a disk, rather than a sphere.

On the other hand, the insolation reduction discussed above was intended to lower the atmospheric equilibrium temperature substantially. Hansen *et al.* [2] have estimated that there is currently a net insolation increase of about 1 W m$^{-2}$ due to greenhouse warming. With the more modest goal of reducing the insolation by this amount, rather than reducing the equilibrium temperature by one Kelvin, we could reduce the cloud optical depth and the needed comet mass by an order of magnitude. Then either the number of tugboats or the time scale could be reduced comparably.

This discussion implies that it is possible that a small comet could be captured into the Earth-Moon system with the aid of moderate thrusts operating for some years, and orbited at several times the Moon's distance from the Earth, for example. However, the transfer from such an orbit to L5 generally would have to be affected on a time scale comparable to a lunar month. It is possible that complex orbits making optimal use of redirection and boost from Earth-Moon fly-bys could accomplish this, but such orbits would be dangerous. It would be more sensible to use smaller telescope tugboats with very focused beams to break up the asteroid into much smaller pieces before attempting this final transfer.

Equation (3) shows that velocity changes of 1 km s$^{-1}$ could be achieved in one month for fragments of radii of a few hundred meters. The additional advantage of working with such small fragments is their large surface area to volume ratio, which would allow them to naturally sublime on a timescale of 50 yrs. This time scale could be greatly shortened by breaking them into still smaller pieces at L5 (and heating them further). This would also be desirable for insuring the Earth's safety.

We conclude that although the tugboat technology has not been tested, and that a number of craft would be needed for the orbit transfers and breakup processes, the production of a dusty moon cloud is conceivable on a useful timescale. This scenario has the advantage that it is scalable. All parts of the scenario could be tested with tiny fragments, e.g., with 'artificial' comets lofted from the Earth or Moon. The dust cloud construction process could be halted or reversed at any point. The cloud would also dissipate naturally over time, unless it was artificially maintained. The diffusive processes discussed above would boost dust grains onto horseshoe orbits from L4 to L5 and around most of the lunar orbit (see Sec. 4). Grains would eventually be boosted to the L1, L2 points near the Moon, and thence into orbits around the Moon, the Earth, or out of the system.

### 3.2   Constructing a Shade with Lunar Material

It is also possible to construct Lagrangian dust shades using lunar material. The possibility of using mass drivers on the lunar surface to boost material to the Lagrange points has been considered in the context of constructing space colonies [10]. Here we merely note a couple of fundamental differences from the use of cometary material. First, the net boost Δ*v* is much less for transport from the lunar surface (only about 1 km s$^{-1}$). However, that velocity change must be achieved impulsively, that is, over a much shorter distance and time. For this reason, the material would have to be transported in many small pieces (boulders), but this is also the case for the final transport phase in the comet scenario. Breaking the boulders down into dust grains would require something like laser vaporization at L5. In the case of the boulders this would require more energy than comet sublimation, but in both cases the energy can be derived directly from the Sun. Research would be needed to determine the best method to break down boulders and achieve a high yield of micron sized dust grains. Additional research is required to show if this is the limiting factor for the use of lunar material.

One could also envision mechanically grinding boulders into dust on the lunar surface and transporting the product in containers. An advantage of this technique is more control over the initial orbital parameters, the importance of which will be discussed in Sec. 4. Lunar mass drivers could accumulate the needed mass by launching about 300 metric tons/s for 10 years.

### 4.   A CLOSER LOOK AT THE EFFECTS OF RADIATION PRESSURE

At the end of Section 2 the timescale for radiation pressure to push grains of the type likely to be obtained from comets was estimated. Since this timescale is not much longer than a lunar month, this effect will be significant, and we will consider it in more detail in this section. Numerical integrations of the equations of motion for a free grain in the Earth-Moon system, with the inclusion of solar radiation pressure, have already been presented by Mignard [14]. However, the examples given were for larger grains than considered above, which according to Eq. (2) have a fairly long timescale for experiencing the effects of radiation pressure. In this section The results of similar integrations are presented, but sampling a somewhat larger range of grain parameters and initial conditions, to get a more complete idea of the persistance of grains at these locations.

We begin with the restricted three-body equations of motion for a free dust particle in the Earth-Moon system, with the x-y plane taken as that of the lunar orbit. We also work in a system rotating with the lunar period, with the now fixed line connecting the Earth and the Moon taken as the x-axis. Finally, the motion in the direction perpendicular to the lunar orbital plane (the z-direction) will be neglected, since we are primarily interested in the escape of dust grains, rather than their detailed distribution around the lunar orbit. Then the equations of motion for the dust grain are,

$$\ddot{x} = 2\omega\dot{y} + \omega^2 x - \frac{Gm_1}{R_1^3}(x-b) - \frac{Gm_2}{R_2^3}(x+a),$$
$$\ddot{y} = 2\omega\dot{x} + \omega^2 y - \left[\frac{Gm_1}{R_1^3} - \frac{Gm_2}{R_2^3}\right] y, \quad (4)$$

where $x$ and $y$ are the particle coordinate positions, subscripts 1 and 2 refer to the Earth and the Moon, respectively, $\omega$ is the orbital frequency, $b$ is the Earth's distance from the center of mass, and $a$ is the Moon's distance (also see Sec. 3.2 of ref. 9). The $R$ terms are give by,

$$R_1 = \left[(x-b)^2 + y^2\right]^{1/2},$$
$$R_2 = \left[(x+a)^2 + y^2\right]^{1/2}. \quad (5)$$

The units adopted are the Earth-Moon separation is 1.0, the Earth's mass is 1.0, and the Moon's orbital period is 1.0. In these units the effective gravitational constant is $G' = 37.58$.

In solar system applications, the radiative acceleration is conventionally given in terms of the solar gravitational acceleration,

$$F_{rad} = \beta \frac{GM_o}{a_o^2} = \frac{16.4}{r_{gr}(\mu m)}, \quad (6)$$

where $M_o$ is the mass of the Sun, and $a_o$ is the distance from the Sun. The second equality makes use of the relation $\beta = 0.2/r_{gr}(\mu m)$, where $r_{gr}(\mu m)$ is the grain radius in microns (see ref. 14 and Sec. 2.7 of ref. 7). This is the magnitude of $F_{rad}$, but in the rotating coordinate system the Sun seems to rotate around the system, so the components of $F_{rad}$ in the equations of motion must include circular functions of $\omega t$. (The Earth-Moon motion around the Sun and the difference between the resulting synodic period and the sidereal period are not included.)

Four dust particle trajectories are shown in Fig. 1. In all cases the particles start on the lower Lagrange point, but with slightly nonzero velocities relative to that point, as indicated in the captions. In all cases the value $\beta = 0.002$ was used. This corresponds to a grain size of 100 $\mu m$ for an unshaded grain, or less for a shaded grain. All four trajectories are calculated for a long enough time to move more than two Earth-Moon separation distances from the Earth. At that distance, radiation pressure generally becomes the dominant force, and in the rotating frame of reference the particle seems to spiral outward. The first two trajectories exceeded this distance in less than 24 lunar months, the third in less than 15 months, and the fourth in less than 18 months.

The first particle starts out on an orbit similar to the quasi-elliptical orbits found near a Lagrange point in the absence of radiation pressure, but is later perturbed into a highly erratic horseshoe-like orbit, and then pushed out. The second particle does one loop on a tadpole-like orbit, and is then pushed out, though it stays in the vicinity for a considerably longer time than most. The third particle stays on a tadpole-like orbit for some months. The fourth remains for several cycles of a horseshoe-like orbit.

In these and many other trajectories most particles are ejected following a close encounter with the Moon, probably on orbits along the so-called interplanetary transport network [21]. A much smaller fraction (less than 10%) ever come close to Earth. Generally, as $\beta$ is increased (smaller grains), the particles are lost more quickly, as expected with greater radiation pressure. However, there is a very strong stochastic element to this process, probably as a result of great sensitivity to the occurrence of lunar encounters. Thus, it is not uncommon for a particle of lower $\beta$ to be lost before a particle of higher $\beta$ with the same initial position and velocity.

There is also sensitivity to the initial solar position or phase relative to the particle velocity. In all the trajectories in Fig. 1, the Sun was initially located on the positive x-axis.

Both of these sensitivities will also remain in the case of partial or periodic shading of the grain in a thick cloud. Investigation of the details of this process is beyond the scope of this paper, but we expect that shading could increase the average retention time of grains.

Another force we have not included is the solar tidal force. Recall that Mignard [14] found that long-lived orbits were still possible in the Earth/Moon system when that force was included.

We have also found that there exist special, very long-lived orbits in the presence of radiation pressure, and Fig. 2 shows two examples for different values of b. In both cases the particles follow quasi-elliptical orbits. The particles persist in these orbits for times in excess of 1000 lunar months or nearly 80 years. Presumably these are resonant orbits, with orbital periods such that the outward push of radiation pressure at one phase is nearly balanced by the inward push at the opposite phase. Note also that taking the endpoints of the trajectories in Fig. 2, and reversing the velocities, should yield trajectories that spiral inward to the Lagrange point, but then leave the system promptly. We can conjecture that resonant orbits would still exist when the effects of solar tides and shading were included, though they would likely be much rarer.

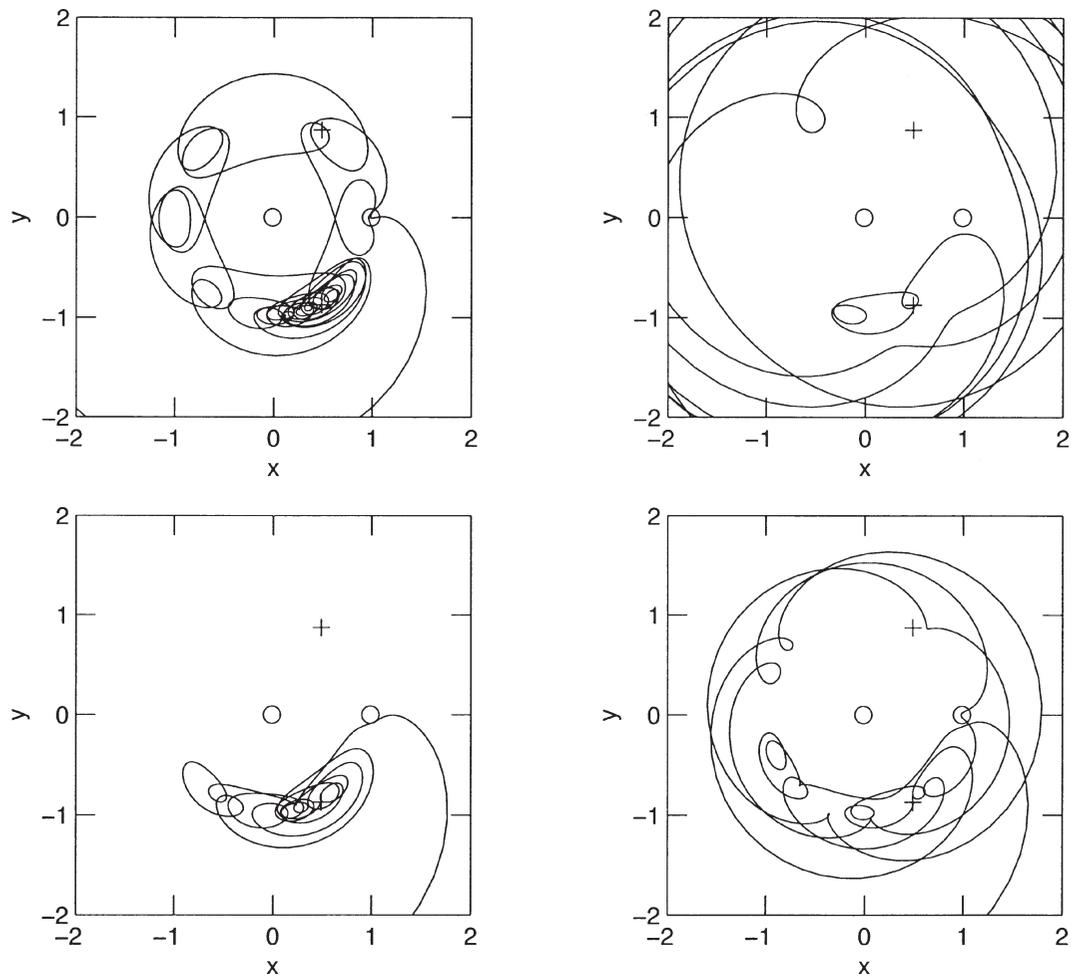

**Fig. 1** Each panel shows a single dust grain trajectory subject to the gravity of the Earth and the Moon, and the solar radiation pressure in a rotating frame. The radiation force is that appropriate to an unshaded grain of radius 100 microns (see text). Each grain begins at the lower Lagrange point, with the Sun located on the positive x-axis, and is followed until its distance from the Earth is more than twice that of the Moon. The $L_{4,5}$ Lagrange points are indicated by crosses. The circle on the left gives the position of the Earth, and the circle on the right gives the position of the Moon. The panels are distinguished by the initial velocity vectors of the grains, $(v_x, v_y)$. The initial velocity coordinates of the grain in: the upper left panel are (-17 m/s, 0.0), the upper right panel are (17 m/s, 0.0), in the lower left panel are (0.0, -17 m/s), and in the lower right panel are (0.0, 17 m/s).

Figure 3 shows the y coordinate (arbitrarily chosen) versus time for the same orbits as shown in Fig. 2. We see that the orbits are nearly resonant, since the time between y maxima is very close to 1.0 lunar period. Because these two periods are not exactly in resonance, a "beating" envelope is evident in Fig. 3. This is interesting because it means that the grain orbit around the Lagrange point cycles between large and small sizes.

In sum, we could expect some grains released from a captured comet to persist in a torus around the lunar orbit for timescales of years, though most of the smaller ones would be blown away more quickly. On the other hand, if the grains could be placed on favorable, resonant orbits, they could be retained for decades in the vicinity of the Lagrange points. Numerical experimentation indicates that there is an elliptical region containing these resonant orbits (or attracting the trajectory to them) in the two-dimensional velocity space, and this region has a mean width of about 0.1 units (or 0.02 km/s) when $\beta = 0.002$. This means that the random, thermal velocity of the grains on these orbits would have to be very small. More realistically, dust grains placed on these orbits would drift away from them, but would still remain in the vicinity of the Lagrange point for much longer times than those placed on random orbits.

These results also have implications for naturally occurring Kordylewski clouds [15, 16]. Specifically, if real, they would most likely consist of quite large grains. Even so, they would not persist for long without sources for new grains. Sporadic collisions between larger rocky bodies might provide such sources, but there would have to be a large population of source bodies to maintain the clouds. It is also possible that some meteoritic material follows escape trajectories like those shown in Fig. 1, but in reverse, resulting in the placement of particles near the L4, 5 points. A quantitative evaluation of this process is beyond the scope of this paper.

## 5. BRIEF COMPARISONS TO OTHER GLOBAL SHADING SCHEMES

We can begin a more general consideration of ways to cut down solar insolation, we might ask – if the material can be derived from the Moon, why not simply create a dust cloud around the Moon? In principle, one could derive some of the same benefits without creating completely new sources of nighttime lighting. (Of course, some material placed at either location will leak to the other, but probably not as a very effective obscuring cloud.) However, because the gravitational potential of the Moon is much steeper than that near the Lagrange points it would be harder to make an extensive obscuring cloud. Material would have to be placed more systematically at different radial distances from the Moon. On the plus side, the material would orbit in a (thin) disk rather than a spherical cloud. On the other hand, relatively strong shear effects in lunar orbit would have to be considered. Moreover, the disk will be perturbed by the solar radiation, and these effects will couple to the shear, introducing more complexity.

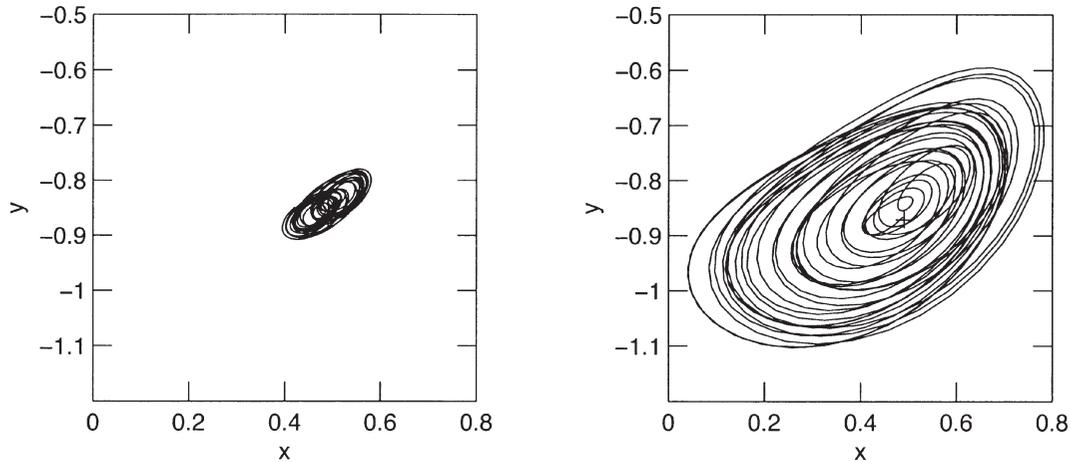

**Fig. 2** Trajectories as in Fig. 1, but with an expanded scale around the initial position at the Lagrange point. Specifically, the upper Lagrange point and the Earth and Moon are not shown. The scale is the same in both panels. In the left panel the unshaded grain size is assumed to be 1000 microns, and in the right panel it assumed to be 100 microns. The initial velocity vector is (-22 m/s, -22 m/s). Only the first 30 lunar months of each trajectory is shown, but these persistent trajectories have been found to stay within the outer positional boundaries shown here for 1000 lunar months.

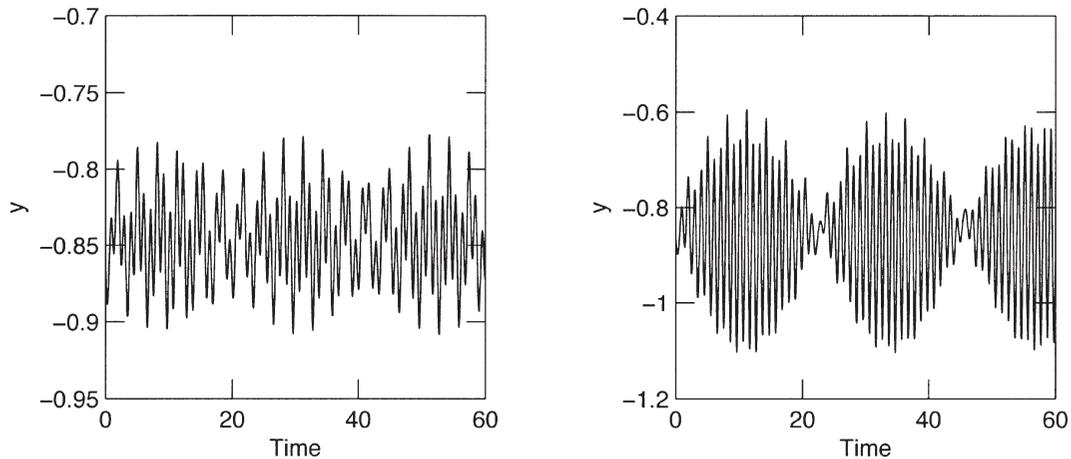

**Fig. 3** The same trajectories as shown in Fig. 2, but here the y-coordinate is shown as a function of time for 60 lunar months.

Perhaps the strongest argument against a lunar cloud is that although there would be continual erosion, its dissipation time would likely be much longer than that of Lagrange clouds. That is, it might be hard to get rid of. The ability to promptly reverse the effects would be very a desirable feature of any planetary scale engineering projects where all the effects can not be predicted in advance.

Next, let us briefly consider some very different types of shading processes. First, we note that ring systems have been considered as a means of reducing solar insolation [4]. For Earth ring systems, it would take a substantially greater $\Delta v$ to get the material into Earth orbit (from either above or below) as compared to the L4, L5 points. Also the sky brightness, and meteor problems described below would be worse. The worst problem with this solution is the hazard to Earth-orbiting spacecraft. It seems that this solution might be more appropriate for moderating Venus' climate than Earth's.

A shield or lens at the Earth/Sun L1 point [3, 4] is in many ways the most elegant space-based solution for reducing solar insolation. It would not brighten the night sky or rain micrometeors on the Earth. The main practical difficulty is that the position and angle of a structure of diameter of about 1000 km must be maintained with great accuracy for a time of order decades. The control equipment and communications lines would also have to be hardened against solar flares. Orbital station-keeping against external perturbations and radiation pressure would require only modest thrusts. Moreover, the mass needed for a disk at L1 is four orders of magnitude less than the comet cloud, so transport would be correspondingly easier.

Any structure that collects or could collect large amounts of energy, or which redirects large fluxes of radiant energy, can be viewed as a potential weapon. This poses a considerable politi-cal barrier to the construction of an L1 lens or shield. Strong safeguards would have to be worked out. Large comet fragments at lunar L5 could not be quickly de-orbited onto Earth with the tugboats described above. Even if they could, their landing site could not be reliably predicted, so they have negligible weapons potential. On the other hand, mass drivers for containers of lunar dust or boulders do have some weapons potential.

The lunar L5 comet cloud scenario has another advantage over its rivals. Not only would it provide some shade from optical radiation, it would also do so for harder radiation from solar flares, and from the particles in coronal mass ejections. It would do so not only for the Earth, but also for objects in near-Earth orbit, and near the L4, L5 points.

## 6. SOME CONSEQUENCES

Planetary scale engineering will involve planetary scale consequences, and there are several quite obvious ones in the dust cloud scheme. One is that the steady meteor flux on Earth will be increased, as will the micro-meteor hazard in Earth and lunar orbits. An accurate estimate of these effects will require detailed simulations, though the total mass is not large. Consider the following simple, but somewhat ad hoc estimate. A $10^{14}$ kg body has the equivalent of about $10^{24}$ grains of radius 20 mm. Suppose 1% or $10^{22}$ grains fall to Earth on a 10 yr. timescale. (In Sec. 4 I estimated that less than 10% of the cloud material would hit Earth, and presumably it would do so on a longer total timescale.) Then the meteor flux is,

$$\frac{10^{22}}{10\, yr. \pi R_E^2} = 1.5 x 10^7 \min^{-1} km^{-2} = 15 \min^{-1} m^{-2}, \quad (7)$$

or about 1 every 4 sec. per m$^2$. This is a large flux, but these meteors would be so small, that they would be very hard to detect. They might affect the physics and chemistry of the upper atmosphere however.

There is also a danger of meter-sized pieces of debris finding their way to Earth. The large centrifugal barrier would prevent that in most cases, and such mini-comets would be easily detectable targets for interception. The regions around the triangular Lagrange points would be polluted by objects covering a large range of sizes, making them less attractive as sites for space colonization, though this is a problem to be faced by any scheme to move large quantities of material there.

The Earth's night sky view would be drastically changed, with very bright moon clouds at L4, L5 and around the lunar orbit. If these clouds were largely translucent their monthly phases would not be as dramatic as the opaque Moon; they would be "full" at all times. With their larger scattering area they would be brighter, and generally the night sky would be much brighter than currently at full moon. Ground and Earth-orbit based astronomy would be devastated in many wavebands. However, brighter nights would reduce the need for excessive artificial lighting.

The effects of the increased night light on the Earth's ecologies would be extremely complex and difficult to predict (see 22). Equally complex would be the effects of driving the annual and long-term climate with periodic changes in solar radiation. The legal, political and economic issues associated with such planetary engineering are also very complex [23].

In fact, almost all aspects of this type of project are complex, with many uncertainties in both the science and the engineering, yet overall, nothing suggests that it would be impossible on a time scale of several decades to a century. More detailed studies and tests could be accomplished on the orbital mechanics, the tugboat spacecraft and the environmental effects in the next ten to twenty years. If these tests were successful and the prospects for global warming remained grim, a full-scale project might be warranted. Humanity has already made changes that adversely affect the global environment of Earth. With the methods described above we may be able to do so in ways that improve that environment.

## ACKNOWLEDGEMENTS

I am grateful to my colleagues Charles Kerton, Bev Smith, and Lee Anne Willson for providing input on early versions of this work.

## REFERENCES

1. Intergovernmental Panel on Climate Change (IPCC), *Climate Change 2001: The Scientific Basis,* eds. J.T. Houghton *et al*., Cambridge Univ. Press, New York, 2001.
2. J. Hansen *et al.*, "Earth's Energy Imbalance: Confirmation and Impliction", *Science*, **308**, pp.1431-1435, 2005.
3. E. Teller, L. Wood and R. Hyde, "Global Warming and Ice Ages: I. Prospects for Physics-Based Modulation of Global Climate Change," from the *22$^{nd}$ International Seminar on Planetary Emergencies*, Sicily, Italy, August 1997. Lawrence Livermore National Laboratory preprint UCRL-JC-128715, 1997.
4. J. Pearson, J. Oldson and E. Levin, "Earth Rings for Planetary Environment Control," paper presented at the *53th International Aeronautical Conference*, IAF-02-U.1.01, 2002
5. A.J. Tuzzolino *et al.*, "Dust Measurements in the Coma of Comet 81P/Wild 2 by the Dust Flux Monitor Instrument," *Science*, **304**, pp.1776-1780, 2004.
6. M.F. A'Hearn *et al.*, "Deep Impact: Excavating Comet Tempel 1", *Science*, **310**, pp.258-264, 2005.
7. I. De Pater and J.J. Lissauer, *Planetary Sciences*, Cambridge Univ. Press, Cambridge, 2001.
8. H.U. Keller *et al.*, "Deep Impact Observations by OSIRIS Onboard the Rosetta Spacecraft", *Science*, **310**, pp.281-283, 2005.
9. C.D. Murray and S.F. Dermott, *Solar System Dynamics*, Ch. 3, Cambridge Univ. Press, Cambridge, 2001.
10. R.D. Johnson and C. Holbrow (eds.), "*Space Settlements A Design Study: NASA SP-413*" Ch. 2, NASA, Washington, 1977.
11. A.A. Kamel and J.V. Breakwell, "Stability of Motion near Sun-Perturbed Earth-Moon Triangular Libration Points," in *Periodic Orbits Stability and Resonances*, ed. G.E.O. Giacaglia, pp.82-90, Reidel, Dordrecht, 1970.
12. B.E. Schutz and B.D. Tapley, "Numerical Studies of Solar Influenced Particle Motion Near the Triangular Earth-Moon Libration Points", in *Periodic Orbits Stability and Resonances*, ed. G.E.O. Giacaglia, pp.128-142, Reidel, Dordrecht, 1970.
13. C. Díez, A. Jorba and C. Simó, "A Dynamical Equivalent to the Equilateral Libration Points of the Earth-Moon System", *Cel. Mech. & Dyn. Astron.,* **50**, pp.13-29, 1991.
14. F. Mignard, "Stability of $L_4$ and $L_5$ Against Radiation Pressure", *Cel. Mech.*, **34**, p.275, 1984.
15. K. Kordylewski, "Photographic Investigation of the Libration Point $L_5$ in the Earth-Moon System", *Acta Astron.*, **11**, p.165, 1961.
16. N.S. Moeed and J.C. Zarnecki, "Feasibility of Space Based Observations of the Kordylewski Clouds", *Adv. Space Res.,* **20**, p.1527, 1997.
17. E.T. Lu and S.G. Love, "A Gravitational Tractor for Towing Asteroids",

*Nature*, **438**, pp.177-178, 2005.
18. H.J. Melosh, I.V. Nemchinov and Y.I. Zetzer, "Non-Nuclear Strategies for Deflecting Comets and Asteroids", in *Hazards due to Comets and Asteroids*, ed. T. Gehrels, pp.1111-1134, Univ. of Arizona Press, Tucson, 1994.
19. P. Venetoklis, E. Gustafson, G. Maise and J. Powell, "Application of Nuclear Propulsion to NEO Interceptors", in *Hazards due to Comets and Asteroids*, ed. T. Gehrels, pp.1089-1110, Univ. of Arizona Press, Tucson, 1994.
20. S.R. Chesley and T.B. Spahr, "Earth Impactors: Orbital Characteristics and Warning Times", in *Mitigation of Hazardous Comets and Asteroids*, eds. M.J.S. Belton *et al*., pp.22-37, Cambridge Univ. Press, Cambridge, 2004.
21. G. Gómez, W.S. Koon, M.W. Lo, J. Marsden, J. Masdemont and S.D. Ross, "Connecting Orbits and Invariant Manifolds in the Spatial Restricted Three-Body Problem", *Nonlinearity,* **17**, p.1571, 2004.
22. C. Rich and T. Longcore (eds.), "*Ecological Consequences of Artificial Night Lighting*", Island Press, Washington, D. C., 2005.
23. E. Fasan, "Planetary protection - some legal questions", *Advan. Space Res.*, **34**, pp.2344-2353, 2004.

\* \* \*

Reference to the work of Katz (1975) was omitted in the previous version. Katz found that including the real eccentricities of the Earth-Moon orbit decreased the time for the loss of test particles placed near L4,5. He also noted the effect of radiation pressure on small grains.

J. I. Katz, "Numerical orbits near the triangular lunar libration points", *Icarus,* **25**, pp.356-359, 1975